\documentclass{elsart3p}             

\usepackage{graphicx}
\usepackage{amsmath,amssymb,amsfonts}
\usepackage{euscript}

\usepackage[pdfpagemode=UseNone,pdfstartview=FitH]{hyperref}

\usepackage{ifthen}
\newboolean{@IsSubmitted}
\setboolean{@IsSubmitted}{false}
\setboolean{@IsSubmitted}{true}      

\newboolean{@IsBWplots}
\setboolean{@IsBWplots}{true}        
\setboolean{@IsBWplots}{false}

\newboolean{@IsPstBarFigs}
\setboolean{@IsPstBarFigs}{false}

\ifthenelse{\boolean{@IsSubmitted}}{ 
  \usepackage{multirow}
  \newcommand{\en}[1]{(\ref{eq:#1})}
  \newcommand{\mymatrix}[1]{\boldsymbol{\mathsf{#1}}}
  \newcommand{\units}[1]{\mbox{$\mathrm{#1}$}}
  \newcommand{\leavethisout}[1]{}
  \newcommand{\mymatrixentry}[1]{\mathsf{#1}}
}{
  \input{mydefs}
}

\graphicspath{{lushi-images/}{lushi-report/}}
\DeclareGraphicsExtensions{.eps,.ps,.pdf,.jpg}

\newcommand{\bvec}[1]{\vec{#1}}
\newcommand{\Wset}{W_{\text{{set}}}}
\newcommand{\Wdep}{W_{\text{{dep}}}}
\newcommand{\Umin}{U_{\text{{min}}}}

\newcommand{\Sheight}{H}
\newcommand{\Rheight}{h}
\newcommand{\zinc}{\text{Zn}}
\newcommand{\strontium}{\text{Sr}}

\newcommand{\sulphate}{\text{SO${}_4$}}
\newcommand{\computername}{MacBook Pro with a 2.5 GHz Intel dual-core processor} 
\newcommand{\myunits}[1]{[\units{#1}]}

\renewcommand{\mymatrix}[1]{\mathbb{#1}}

\newcommand{\mycitename}[2]{#2\ \cite{#1}}
\newcommand{\mycite}[1]{\cite{#1}}

\begin{document}

\begin{frontmatter}

  \title{An inverse Gaussian plume approach for estimating atmospheric
    pollutant emissions from multiple point sources
}
  
\author[lushi]{Enkeleida Lushi} \and
\ead{lushi@cims.nyu.edu}
\author[stockie]{John M. Stockie\corauthref{jscor}}
\corauth[jscor]{\raggedright Corresponding author.  Tel.:~+1~778~782~3553,  
  Fax:~+1~778~782~4947.} 
\ead{stockie@math.sfu.ca}

\address[lushi]{Courant Institute of Mathematical Sciences, New York
  University, 251 Mercer Street, New York, NY, 10012, USA}
\address[stockie]{Department of Mathematics, Simon Fraser University,
  8888 University Drive, Burnaby, BC, V5A 1S6, Canada}

\begin{abstract}
  A method is developed for estimating the emission rates of
  contaminants into the atmosphere from multiple point sources using
  measurements of particulate material deposited at ground level.  The
  approach is based on a Gaussian plume type solution for the
  advection--diffusion equation with ground-level deposition and given
  emission sources.  This solution to the forward problem is
  incorporated into an inverse algorithm for estimating the emission
  rates by means of a linear least squares approach.  The results are
  validated using measured deposition and meteorological data from a
  large lead-zinc smelting operation in Trail, British Columbia.
  The algorithm is demonstrated to be robust and capable of generating
  reasonably accurate estimates of total contaminant emissions over the
  relatively short distances of interest in this study.
\end{abstract}


\end{frontmatter}

\section{Introduction}
\label{sec:intro}

Urban air quality is an issue of major concern owing to recent upward
trends in population growth and urbanisation and industrialisation
around the world.  Consequently, there is an increasing need to
understand the detailed dynamics governing emission and transport of
particulate matter in the atmosphere.  \mycitename{turner-1979}{Turner}
mentions a multitude of possible sources of air-borne particles,
including those of anthropogenic origin such as industrial complexes and
automobiles, as well as natural sources such as dust storms and volcanic
eruptions.  Recently, there has been a surge of interest in related
problems for disaster planning and national security that involve
transport of radionucleotides and biological or chemical agents
\mycite{settles-2006}.

The physics of particulate transport in the atmosphere are complex, in
many cases involving multiple spatial scales (ranging from the particle
scale to near-source and long-range effects), multi-physics (coupling
mass transport, turbulence, chemistry and wet/dry deposition), and
complex geometry (e.g., involving flow over topography or man-made
structures).  Models for these and other aspects of atmospheric
dispersion have a long history dating back to the pioneering studies of
turbulent diffusion by \mycitename{richardson-1921}{Richardson} and
\mycitename{taylor-1922}{Taylor}.\
The bulk of previous work has tackled the {\itshape forward problem}, by
which we refer to the process of determining downwind contaminant
concentrations given source emission rates and meteorological
conditions.  These forward models are usually based on a solution of the
advection--diffusion equation that is obtained through either analytical
or numerical means (or a combination of both).  The most prevalent
approach used in practice, and which is implemented in many
industry-standard software packages, employs an approximate analytical
solution for point source emissions known as the ``Gaussian plume
solution.'' The one-dimensional plume solution for a single point source
was originally derived by Sutton \mycite{sutton-1932} and has since been
extended to higher dimensions, as well as being applied to a variety of
other more general situations involving ground-level
deposition \mycite{ermak-1977}, multiple sources \mycite{calder-1977},
height-dependent wind speed and diffusion
coefficients \mycite{liley-1995,lin-hildemann-1997}, and line and area
sources, to name just a few.  We remark that while a majority of the
applications considered to date have involved transport in the
atmosphere, the Gaussian plume models just mentioned may also be used
to solve other advection--diffusion problems in such diverse areas as
population growth \mycite{condie-bormans-1997} or\
water flow in rivers \mycite{elbadia-etal-2005} and the
subsurface \mycite{kennedy-ericsson-wong-2005}. 

Another related stream of research has focused on solving the
corresponding {\itshape inverse problem}, whereby measurements of
particulate concentrations or ground-level depositions are given and the
aim is to determine information about the location or efflux rate of
contaminant sources.  Inverse methods based on Gaussian plume type
solutions have been developed by a number of authors in this context
including \mycitename{jeong-etal-1995}{Jeong \etal} and
\mycitename{hogan-etal-2005}{Hogan \etal}, while
\mycitename{mackay-mckee-mulholland-2006}{MacKay \etal} developed an
alternate solution approach using complex variable theory.  Other
researchers have applied a more direct computational approach by solving
the nonlinear governing equations using methods based on Kalman
filtering \mycite{mulholland-seinfeld-1995}, Lagrangian
particles \mycite{seibert-frank-2004}, Bayesian
techniques \mycite{enting-2002,goyal-etal-2005}, or by integrating the
equations backward in
time \mycite{seibert-frank-2004,bagtzoglou-baun-2005}.\ \ 
Several related methods have been developed specifically for handling
the added nonlinearities arising from atmospheric chemistry, typically
using Newton type iterative methods \mycite{brown-1993}, and sometimes
combined with statistical techniques \mycite{houweling-etal-1999}.  These
direct numerical approaches can be very computationally intensive,
especially for 3D problems, and so will typically require use of
parallel computing resources.  Regardless of the numerical method used,
the inverse problem is often characterized as ill-conditioned in the
sense that small changes in parameters can lead to very large changes in
emission estimates; these issues are discussed in much more detail
by \mycitename{enting-2002}{Enting}, \mycitename{beychok-1999}{Beychok},
and \mycitename{elbadia-etal-2005}{El Badia \etal}.

The subject of this paper is a lead-zinc smelting operation located in
Trail, British Columbia, Canada.  We are concerned with the transport of
several contaminant species from multiple point sources on the site, and
our aim is to develop an inverse algorithm that will determine emissions
based on the Gaussian plume solution of \mycitename{ermak-1977}{Ermak}.
The algorithm should be capable of generating reliable estimates of
emissions rates given a relatively small number of deposition
measurements.  The novelty of this work stems from a combination of
factors:
\begin{itemize}
\item We make use of real (noisy) meteorological and deposition data,
  in contrast with some other studies that use synthetic
  data \mycite{hogan-etal-2005,mackay-mckee-mulholland-2006}. 
\item Deposition measurements are relatively small in number and
  represent time-averaged accumulations.  This should be compared with
  some other methods that obtain very high accuracy by using
  very large numbers of sample points \mycite{jeong-etal-1995}.  Another
  example is the work of \mycitename{hogan-etal-2005}{Hogan \etal}, who
  proposed an iterative method based on the Gaussian plume solution
  (with constant wind and no deposition) and which exploits the fact that
  concentration measurements at four locations uniquely determine the 
  location and strength of a single point source.  This approach can be
  very effective when the input data are known very accurately, but it
  degrades when the data are noisy.
\item The emission sources are at known locations, in comparison with
  some other studies that aim to determine both emission rates and
  locations \mycite{mulholland-seinfeld-1995,bagtzoglou-baun-2005,brown-1993}. 
\item We incorporate additional linear constraints on emission rates
  that are derived from chemical processes within the smelting
  operation.
\item Deposition measurements are taken near ground level and at short
  distances from the source, which allows us to avoid errors inherent in
  long-range dispersion estimation and thereby minimize the
  ill-conditioning of the inverse problem.
\end{itemize}
Taken together, these factors allow us to develop a robust algorithm
that is capable of estimating emission sources with a reasonable degree
of accuracy.  Other studies have been performed on emissions at the
Trail site by Goodarzi \etal (see \mycite{goodarzi-etal-2002b}
and references therein), but they use a much simpler Gaussian plume
solution with no deposition and constant wind velocity, as well as
validating their results using long-range deposition measurements and
different experimental techniques.

We begin in Section \ref{sec:problem} by describing the problem under
study and developing a detailed list of assumptions underlying the
model.  In Section~\ref{sec:plume-model} we provide details of the Ermak
solution to the advection--diffusion equation, and also incorporate
multiple sources and a time-varying wind velocity.
Section~\ref{sec:inverse} focuses on the inverse problem, deriving the
linear equality and inequality constraints and describing the linear
least squares solution algorithm.  A series of numerical simulations are
performed in Section~\ref{sec:sims}, including a study of the
sensitivity of the model to changes in parameters and noise in the data.
Finally, we conclude with recommendations about the suitability of
applying our model to actual environment reporting scenarios, and make
suggestions on possible future work on extending our approach in order
to improve accuracy and permit application to a wider range of
atmospheric dispersion scenarios.

\section{Problem description and simplifying assumptions}
\label{sec:problem}

The motivation for this work was a study of emissions from a number of
contaminant sources at a large lead-zinc smelter located in Trail,
British Columbia, Canada and operated by Teck Cominco Limited.  Our
primary aim was to improve the accuracy of airborne emission estimates
(especially for zinc) that the Company is required to report annually to
Environment Canada's {\itshape National Pollutant Release Inventory
  (NPRI)} \mycite{npri}.  Stack emissions on the smelter site
are directly measured for zinc and other contaminants, but this paper
deals with low-level sources for which direct measurements are
difficult to obtain.

\begin{figure*}[tbhp]
  \centering
  \ifthenelse{\boolean{@IsSubmitted}}{ 
    \includegraphics[width=\textwidth]{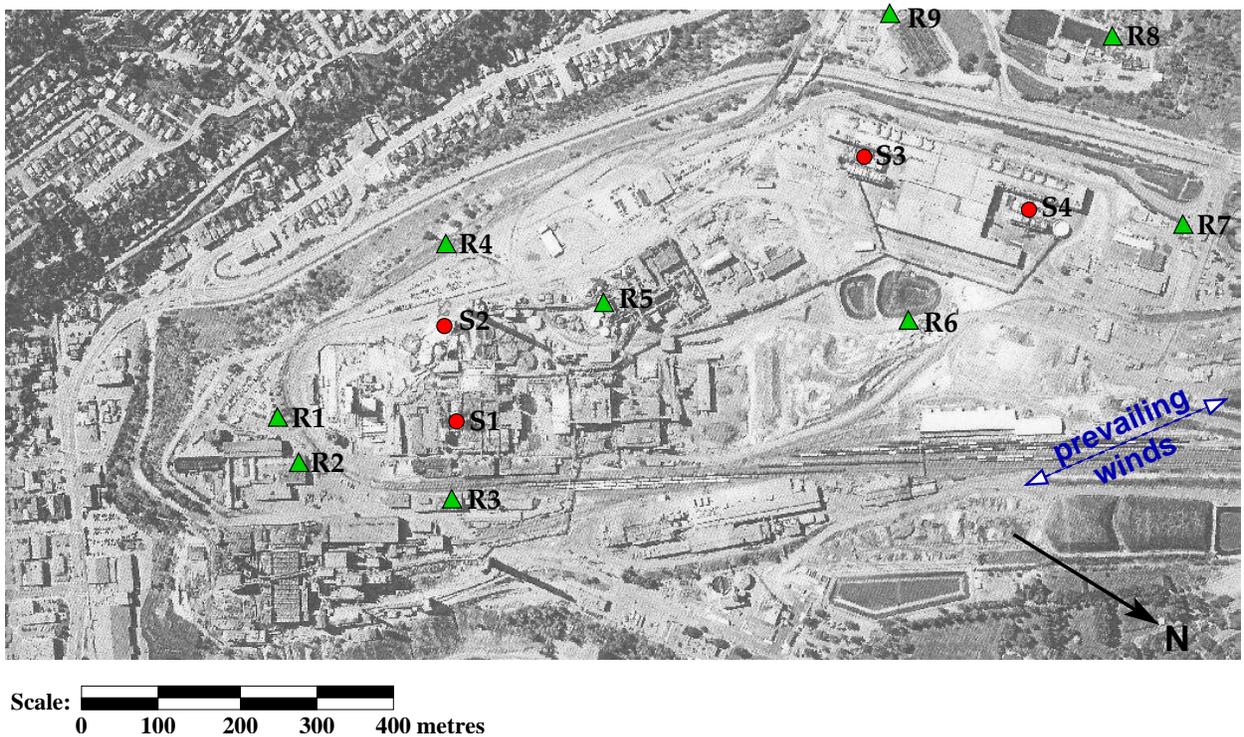}      
  }{
    \includegraphics[width=\textwidth]{trailsite3}   
  }
  \caption{An aerial photo of the Trail site, showing the approximate
    locations of each of the sources (red circles, labeled by S$s$,
    $=1,2,3,4$) and receptors (green triangles, labeled R$r$,
    $r=1,2,\dots, 9$).  The size of the area depicted here is
    approximately $1600 \times 800$~\units{m} and the directions of both
    the prevailing winds and compass north are indicated in the lower
    right corner. 
  }
  \label{fig:trailsite}
\end{figure*}

There are four sources on the Trail site that emit zinc (in the form of
zinc sulphate, \zinc\sulphate) and these are indicated on the aerial
photo in Fig.~\ref{fig:trailsite} by the symbol S$s$, where $s=1,2,3,4$.
To assist in estimating the level of zinc emissions, the Company
performed a series of ground-level measurements of zinc as well as a
number of other contaminant species (strontium, sulphur, etc.).  The
measurements were taken over the two-year span 2001--2002 and consist of
one-month accumulations of particulates within dustfall jars or
``receptors,'' which are located at nine separate locations R$r$,
$r=1,2,\dots, 9$ (also indicated in Fig.~\ref{fig:trailsite}).

Meteorological data is available for the same monthly periods in terms
of wind speed and direction averaged over 10-minute intervals.  The
smelter is located in the Columbia River valley which tends to funnel
the winds on the site in a specific direction; since the river is
located just below the aerial photo in Fig.~\ref{fig:trailsite} and runs
roughly horizontally, the prevailing wind directions are roughly to the
northwest or southeast.

Based on the above description, we now make a number of assumptions
which are critical in the development of our dispersion model:
\begin{enumerate}
\item Each emission source is considered to be a point source,
  and all emission rates are constant in time, at least for any
  one-month period over which depositions are measured.
\item The\label{assume:wind} wind velocity $\bvec{u}(t)$ depends on time
  only and is uniform throughout the domain at any given instant.  This
  is justifiable considering the relatively small dimensions of the site
  (1600 $\times$ 800 \units{m}) and the short time intervals of
  interest.
\item Variations in topography are negligible so that the wind can be
  assumed horizontal.  Although Trail is located in a river valley
  bordered by steep mountains, the domain of interest is far enough
  removed from the neighbouring mountain range that any topography or
  boundary effects can be ignored.
\item A ten-minute averaging period (or time step) is used in all
  calculations, which is consistent with the assumptions employed in
  deriving the Gaussian plume
  solution \mycite{beychok-1999,hanna-briggs-hosker-1982} to limit
  errors in concentration.
\item Only dry deposition is considered since the dustfall measurements
  were all taken during months for which rainfall is relatively small.
\item The effects of plume rise are incorporated by using an ``effective
  height'' for each stack.
\item The at\-mo\-spher\-ic sta\-bil\-i\-ty class is assumed to be class
  D (``neutral'', according to the Pasquill-Gifford classification
  scheme) for all monthly periods.  There are insufficient
  meteorological data available to consider varying the stability class
  with time, and so our choice of neutral class is a compromise that
  takes into account predominant atmospheric conditions during the
  months of interest.
\end{enumerate}
It is important to note that even though we focus our attention on
an application to a specific industrial site, the mathematical model and
associated numerical algorithms we develop are general and so can
potentially be applied to a wide range of other atmospheric dispersion
problems.

\section{Derivation of the forward model}
\label{sec:plume-model}

 
The release and transport of a single contaminant in the atmosphere can
be described by the advection--diffusion equation
\begin{gather}
  c_t + \bvec{u} \cdot \nabla c = \nabla \cdot ({\cal K} \nabla c) 
  + S,
  \label{eq:advdiff}
\end{gather}
where 
\begin{center}
  \begin{tabular}{rp{0.75\columnwidth}}
    $c(\bvec{x},t)=$ & contaminant concentration \myunits{kg\,m^{-3}}, \\
    $\bvec{u}=$      & convective wind velocity \myunits{m\,s^{-1}}, \\
    ${\cal K}(\bvec{x})=$ & $\operatorname{diag}(K_x,K_y,K_z)$, the matrix
    of turbulent eddy diffusivities \myunits{m^2\,s^{-1}},\\
    $S(\bvec{x},t)=$ & emission source term \myunits{kg\,m^{-3}\,s^{-1}}.\\
  \end{tabular}
\end{center}
We begin by considering a single elevated point source at location
$(0,0,\Sheight)$ for which the source term takes the form
\begin{gather}
  S(\bvec{x},t) = Q \, \delta(x) \, \delta(y) \, \delta(z-\Sheight), 
\end{gather}
where $Q$ is a constant emission rate \myunits{kg\,s^{-1}} and $\delta$
represents the Dirac delta function \myunits{m^{-1}}.\ \ 
The coordinates $\bvec{x}=(x,y,z)$ are chosen so that the $x$-axis is
aligned parallel to the wind velocity.  Although the wind velocity lies
within the horizontal plane by our earlier assumption
(\ref{assume:wind}), there is nonetheless a small vertical velocity
component $\Wset$ that corresponds to gravitational settling.  As a
result, the velocity appearing in \en{advdiff} takes the form
$\bvec{u}=(U(t),0,-\Wset)$, where $U(t)$ is the wind speed and the
constant settling velocity is determined for spherical particles using
Stokes' law
\begin{gather}
  \Wset = {\rho g d^2}/{(18\mu)},
  \label{eq:stokes}
\end{gather}
where
\begin{center}
\begin{tabular}{rp{0.78\columnwidth}}
  $\rho=$ & particle density \myunits{kg\,m^{-3}},\\
  $d=$    & particle diameter \myunits{m}, \\
  $\mu=$  & air viscosity $=1.8\times 10^{-5}$ \myunits{kg\,m^{-1}\,s^{-1}},\\
  $g=$    & gravitational acceleration $=9.8$ \myunits{m\,s^{-2}}.
\end{tabular}
\end{center}

Diffusive transport in the $x$-direction is typically much smaller than
convective transport by the wind; hence, the diffusion term involving
$K_x$ may be neglected, and $K_y(x)$ and $K_z(x)$ may be taken as
functions of downwind distance only.  As a result, Eq.~\en{advdiff}
reduces to 
\begin{multline}
  \frac{\partial c}{\partial t} + U(t) \, \frac{\partial c}{\partial x} 
  - \Wset \,\frac{\partial c}{\partial z} = \\
  \frac{\partial}{\partial y} \left( K_y \frac{\partial c}{\partial y}\right) 
  + \frac{\partial}{\partial z} \left( K_z \frac{\partial c}{\partial z}\right) 
  + S(\vec{x},t)
  .  \label{eq:advdiff2} 
\end{multline}
This equation could be modified to include vertical variations in both
wind velocity and dispersion coefficients following
\mycite{chrysikopoulos-hildemann-roberts-1992,koch-1989}, although such
effects will not be considered in this paper.

For the moment, we assume the domain is of infinite extent in the $x$ and
$y$ directions, and the positive $z$ direction; consequently,
we can apply the following Dirichlet boundary conditions at infinity:
\begin{align}
  c(x,\pm \infty, z) &= 0, \label{eq:bc1}\\
  c(\pm \infty,y, z) &= 0, \label{eq:bc2}\\
  c(x,y,\infty)      &= 0. \label{eq:bc3}
\end{align}
The ground surface ($z=0$) is where deposition occurs, and so we impose
the following mixed (Robin type) condition on particle flux
\begin{gather}
  \left.\left(K_z \frac{\partial c}{\partial z} + \Wset\, c \right)
  \right|_{z=0} = \left. \Wdep\, c \right|_{z=0},
  \label{eq:bcdep}
\end{gather}
where $\Wdep>0$ is the dry deposition velocity which is usually assumed
constant and is determined from experiments.

\subsection{Ermak's solution}
\label{sec:plume-eqns}

The Eqs.~\en{advdiff2}--\en{bcdep} have been well-studied and
\mycitename{ermak-1977}{Ermak} derived the following analytical solution
which is valid for $x>0$: 
\leavethisout{
  \begin{align}
    c(x,y,z) = &\,\frac{Q}{2\pi U \sigma_y \sigma_z}
    \exp\left(-\frac{y^2}{2\sigma_y^2} \right)\;
    \exp\left( -\frac{\Wset\,(z-\Sheight)}{2K_z} - \frac{\Wset^2\,\sigma_z^2}{8K_z^2}
    \right)\notag \\
    &\times \left[ \exp\left( -\frac{(z-\Sheight)^2}{2\sigma_z^2} \right)
      + \exp\left(-\frac{(z+\Sheight)^2}{2\sigma_z^2} \right)\right. \notag\\
    &\qquad\left.-\; \sqrt{2\pi}\,\frac{W_o\,\sigma_z}{K_z} \;
      \exp\left( \frac{W_o(z+\Sheight)}{K_z} + \frac{W_o^2\,\sigma_z^2}{2K_z^2}\right)\;
      \operatorname{erfc}\left( \frac{W_o\,\sigma_z}{\sqrt{2}\, K_z} +
        \frac{z+\Sheight}{\sqrt{2}\,\sigma_z}\right) \right],
    \label{eq:c}
  \end{align}
}
\begin{multline}
   c(x,y,z) = \frac{Q}{2\pi U \sigma_y \sigma_z}
  \exp\left(-\frac{y^2}{2\sigma_y^2} \right) \\
   \times
  \exp\left( -\frac{\Wset\,(z-\Sheight)}{2K_z} - \frac{\Wset^2\,\sigma_z^2}{8K_z^2}
  \right)\\
   \times \left[ \exp\left( -\frac{(z-\Sheight)^2}{2\sigma_z^2} \right)
    + \exp\left(-\frac{(z+\Sheight)^2}{2\sigma_z^2} \right)\right. \\
   \left.-\; \sqrt{2\pi}\,\frac{W_o\,\sigma_z}{K_z} \;
    \exp\left( \frac{W_o(z+\Sheight)}{K_z} +
      \frac{W_o^2\,\sigma_z^2}{2K_z^2}\right) \right.\\
    \times \left.\operatorname{erfc}\left( \frac{W_o\,\sigma_z}{\sqrt{2}\, K_z} +
      \frac{z+\Sheight}{\sqrt{2}\,\sigma_z}\right) \right],
  \label{eq:c}
\end{multline}
where $W_o=\Wdep-\frac{1}{2}\Wset$.  The quantities $\sigma_{y,z}(x)$
(having units of \units{m}) represent standard deviations of 
concentration in the $y$ and $z$ directions and can be expressed in
terms of the eddy diffusivities as
\begin{gather*}
  \sigma_{y,z}^2(x) = \frac{2}{U}\int_0^x K_{y,z}(x^\prime) \, dx^\prime.
\end{gather*}
This integral simplifies to $K=\sigma^2U/2x$ when the eddy diffusivities are
constant, although in practice the $\sigma$ values need to be taken as
functions of the downwind distance $x$ in order to match observed plume
concentrations.

Suppose that the height of a receptor above the ground is
$z=\Rheight$; then the key quantity of interest is the contaminant
deposition flux at $z=\Rheight$ which is given by the expression $\Wdep
c|_{z=h}$ from Eq.~\en{bcdep}, with the height replaced by $z=\Rheight$.
It is essential to observe that the deposition flux depends {\itshape
  linearly} on the emission rate $Q$ through Eq.~\en{c}, a fact that
will be exploited to great advantage in this work.

\subsection{Multiple sources and receptors}

Instead of just a single point source, suppose instead that we have a
set of $N_s$ sources, each with emission rate $Q_s$
\myunits{kg\,s^{-1}} and location $\bvec{\xi}_s$, for $s=1,2,\dots,
N_s$.  Likewise, there are $N_r$ receptors located at positions
$\bvec{\eta}_r$, each with total accumulated deposition $D_r$
\myunits{kg}, for $r=1,2,\dots, N_r$.  On the Trail site depicted in
Figure~\ref{fig:trailsite}, there are $N_s=4$ sources and $N_r=9$
receptors that are of interest to this study.  In the derivation that
follows, we note that there are many similarities with
\mycitename{calder-1977}{Calder's} general model framework for
multiple-source emissions, except that he was unconcerned with
deposition or any specific Gaussian plume approximation for the
advection--diffusion equation.

\begin{figure}[tbhp]
  \centering
  \ifthenelse{\boolean{@IsSubmitted}}{ 
    \includegraphics[width=0.8\columnwidth]{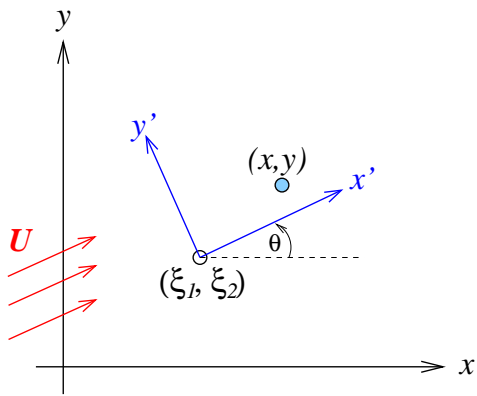}
  }{
    \includegraphics[width=0.8\columnwidth]{coord}
  }
  \caption{Relationship between the original coordinates $\bvec{x}$ and
    transformed coordinates $\bvec{x}^{\,\prime}$ for a given wind field
    having magnitude $U$ and direction angle $\theta$.}
  \label{fig:one}
\end{figure}

In order to apply the solution from \en{c} for a uni-directional wind, a
new set of transformed coordinates $\bvec{\xi}_s$ must be defined for
each source $s$ that translate the source location to the origin and
then rotate coordinates so that the transformed $x$-axis is aligned
parallel with the wind velocity.  To this end, we define new coordinates
$\bvec{x}_s^{\,\prime}$ which are related to $\bvec{x}$ via
\begin{gather}
  \bvec{x}_s^{\,\prime} = {\cal R}_{-\theta} (\bvec{x}-\bvec{\xi}_s),
  \label{eq:coords}
\end{gather}
where $\theta$ corresponds to the angle the wind direction vector makes
with the $x$-axis (measured counter-clockwise) and ${\cal R}_{-\theta}$
represents the matrix that rotates vectors through an angle
$-\theta$ in the $x,y$-plane (see Figure~\ref{fig:one}). The resulting
coordinates have an $x^\prime_s$--axis that is aligned with the wind
direction.
It is convenient to rewrite the concentration from \en{c} by factoring
out the source emission rate and making the dependence on the source
location and wind explicit, letting $c(\bvec{x}_s^{\,\prime})=Q_s\,
p(\bvec{x}; \bvec{\xi_s},\theta,U)$.  Consequently, the deposition flux
may be written as
\begin{gather*}
  \Wdep \, c|_{z=\Rheight} = 
  \Wdep \, Q_s \,p(\bvec{x};\bvec{\xi}_s,\theta,U)|_{z=\Rheight},
\end{gather*}
and we may then calculate the total accumulation of contaminant in
\units{kg} within a dustfall jar at location $\bvec{x}$ over a time
interval of length $\Delta t$ as
\begin{gather*}
  D(\bvec{x}_s^{\,\prime}) = \sum_{s=1}^{N_s} ( \Wdep \,Q_s A \Delta t ) \,
  p(\bvec{x};\bvec{\xi}_s,\theta,U), 
\end{gather*}
where $A$ is the cross-sectional area $A$ of the jar opening.  With a
slight change in notation, we can write the total mass of a single
contaminant deposited at receptor location $\bvec{\eta}_r$ as
\begin{gather}
  D_r = \Wdep A \Delta t \sum_{s=1}^{N_s} Q_s \,
    p(\bvec{\eta}_r;\bvec{\xi}_s,\theta,U)|_{z=\Rheight_r}
  \label{eq:psys}
\end{gather}
which clearly indicates that the deposition is a linear combination of
the emission rates $Q_s$.

\subsection{Time-varying wind}
\label{sec:varwind}

Eqs.~\en{psys} are derived for a wind that is constant over the time
interval of length $\Delta t$.  In practice, the wind velocity varies
with time and these variations have a significant impact on the plume
dispersion and consequently also the deposition.  The wind speed and
direction, $U(t)$ and $\theta(t)$, are therefore both functions of time
and so the coordinate transform \en{coords} is also time-dependent.  In
order to make use of the Ermak solution, we divide time into $N_t$
equally-spaced intervals of length $\Delta t =T/N_t$, and assume that
$U(t)$ and $\theta(t)$ can both be approximated by piecewise constant
functions on each interval.  The effects of time variation may then be
incorporated by solving a sequence of Gaussian plume problems of the
form~\en{c} and the total mass deposited at receptor $r$ is calculated
by summing the results over each time interval,
\begin{gather}
  D_r^{tot} = \Wdep A \Delta t \sum_{s=1}^{N_s} Q_s \sum_{n=1}^{N_t} 
  p(\bvec{\eta}_r;\bvec{\xi}_s,\theta^n,U^n)|_{z=\Rheight_r},
  \label{eq:psys-tot}
\end{gather}
where $p(\bvec{x};\bvec{\xi}_s, \theta^n,U^n)$ represents the expression
for the concentration, using averaged values of $U^n$ and $\theta^n$
corresponding to the $n^{th}$ time interval.

The Gaussian plume model assumes steady-state conditions, and so it is
important to emphasize that we are making a major assumption when
applying it to problems in which the wind velocity varies with time.  As
a result, the time step $\Delta t$ is a critical parameter: it must be
chosen large enough that the plume can be considered sufficiently
developed that it has reached a steady state, and yet not too large that
the wind is seriously under-resolved.  We choose a value of $\Delta
t=10$ \units{min}, motivated by a number of considerations:
\begin{itemize}
\item Since ``typical'' wind speeds lie between 1 and 5
  \units{m\,s^{-1}}, a given plume will be advected over a distance of
  600 to 3000~\units{m}, which is close to the 1000~\units{m} maximum
  separation between any individual source--receptor pair.
\item \mycitename{beychok-1999}{Beychok} advocates an averaging interval
  of 10 minutes or less for the Gaussian plume model based on the
  observation that one-hour intervals can lead to over-prediction of the
  concentration by as much as a factor 2.5.  
\item \mycitename{hanna-briggs-hosker-1982}{Hanna \etal} note that
  Ermak's solution is derived based on the assumption that all
  variables are averaged over time periods of approximately
  $10$~\units{min} in length.
\end{itemize}
Wind data is available down to a resolution of minutes, and so 
presents no limitation on the choice of $\Delta t$.

\subsection{Parameter values and wind data}
\label{sec:params}

The physical parameters corresponding to the contaminant particles are
summarized in Table~\ref{tab:param2}.  Because both zinc and strontium
are deposited as sulphates, the parameters actually correspond to
\zinc\sulphate\ and \strontium\sulphate.  While zinc is the primary
element of interest in this study, we will see in
Section~\ref{sec:inverse} that strontium is an important tracer element
that plays a useful role in determining the solution to the inverse
problem.  The values for particle diameter and deposition velocity are
consistent with data from
\mycite{gatz-1975,mcmahon-denison-1979,pacyna-etal-1989}, while the
settling velocity is computed from Stokes' law \en{stokes}.
\begin{table*}[tbhp]
  \centering
  \caption{Values of the physical parameters for the two
    contaminants of interest, where zinc and strontium actually
    appear in the form of sulphates -- \zinc\sulphate\ and
    \strontium\sulphate.  Note that the particle diameter and
    deposition velocity for \strontium\ (marked $\ast$) have been set
    equal to those for \zinc, in the absence of other more reliable
    estimates.} 
  \begin{tabular}{|ccc|ccccc|}\hline
    {\bfseries Parameter} & {\bfseries Symbol} & {\bfseries Units} &
    \multicolumn{5}{c|}{{\bfseries Value for species $q=$}} \\
    & & & & $\mathbf{\zinc}$ & & $\mathbf{\strontium}$ &
    \\\hline\hline 
    Density   & $\rho^q$ & \hspace*{0.3cm}\units{kg\,m^{-3}}\hspace*{0.3cm} 
    & & 3540 & & 3960 & \\
    Molar mass& $M^q$    & \units{kg\,mol^{-3}}  & & 0.161 & & 0.184 & \\
    Diameter  & $d^q$    & \units{\mu m} & & 5 & & 5${}^\ast$ & \\
    \hspace*{0.3cm}Deposition velocity& $\Wdep^q$ & \units{m\,s^{-1}} 
    & & 0.005 & & 0.005${}^\ast$ & \\
    Settling velocity    & $\Wset^q$ & \units{m\,s^{-1}} 
    & & 0.0027 & & 0.0030 & \\[-0.2cm]
    (from Eq.~\en{stokes}) & & & \hspace*{0.5cm} & & \hspace*{0.5cm} & &
    \hspace*{0.5cm} \\\hline
  \end{tabular}
  \leavethisout{
    \begin{tabular}{|ccc|ZZZZ|}\hline
      {\bfseries Parameter} & {\bfseries Symbol} & {\bfseries Units} &
      \multicolumn{4}{c|}{{\bfseries Value for species $q=$}} \\\cline{4-6}
      & & & $\mathbf{\zinc}$  
      & $\mathbf{\sulphate}$ 
      & $\mathbf{\strontium}$ & \\\hline\hline
      Density   & $\rho^q$ & \hspace*{0.3cm}\units{kg\,m^{-3}}\hspace*{0.3cm} 
      & 7140 & 2070 & 2640 & \\
      Molar mass& $M^q$    & \units{kg\,mol^{-3}}  & 0.065 & 0.096 & 0.088 & \\
      Diameter  & $d^q$    & \units{\mu m} & 0.9 & 0.9${}^\ast$ & 0.9${}^\ast$ & \\
      \hspace*{0.3cm}Deposition velocity& $\Wdep^q$ & \units{m\,s^{-1}} 
      & 0.0062 & 0.0080 & 0.030 & \\
      Settling velocity    & $\Wset^q$ & \units{m\,s^{-1}} & $1.7\times 10^{-4}$ & 
      $5.1\times 10^{-5}$ & $6.4\times 10^{-5}$ & \\[-0.2cm]
      (from Eq.~\en{stokes}) &&&&&&\\\hline
    \end{tabular}
  }
  \label{tab:param2}
\end{table*}

The heights of the four contaminant sources, corrected for plume rise,
are $\Sheight_s=[15,35,15,15]$ while the nine receptors are located at
heights $\Rheight_r=[0, 10, 10, 1, 15, 2, 3, 12, 12]$.  The dustfall
jars are glass containers in the shape of circular cylinders having a
diameter of 0.162~\units{m}, and so the area parameter used in the
deposition calculation in Eq.~\en{psys-tot} is $A=0.0206\;\units{m^2}$.

The other key parameters appearing in the model are the standard
deviations $\sigma_{y,z}$ which we assume take the form $\sigma(x) =
{ax}{(1+bx)^{-c}}$ customarily attributed to
\mycitename{briggs-1973}{Briggs}.  The constants appearing in this
expression depend on the atmospheric stability class, and if we assume
class D as discussed earlier then the values of the parameters are
\mycite{carrascal-etal-1993}:
\begin{align*}
  \text{for $\sigma_y$:} & \quad a=0.08,\;\; b=0.0001,\;\; c=0.5,\\
  \text{for $\sigma_z$:} & \quad a=0.06,\;\; b=0.0015,\;\; c=0.5.
\end{align*}

\begin{figure}[tbhp]
  \centering
  \ifthenelse{\boolean{@IsSubmitted}}{ 
    \includegraphics[width=0.8\columnwidth]{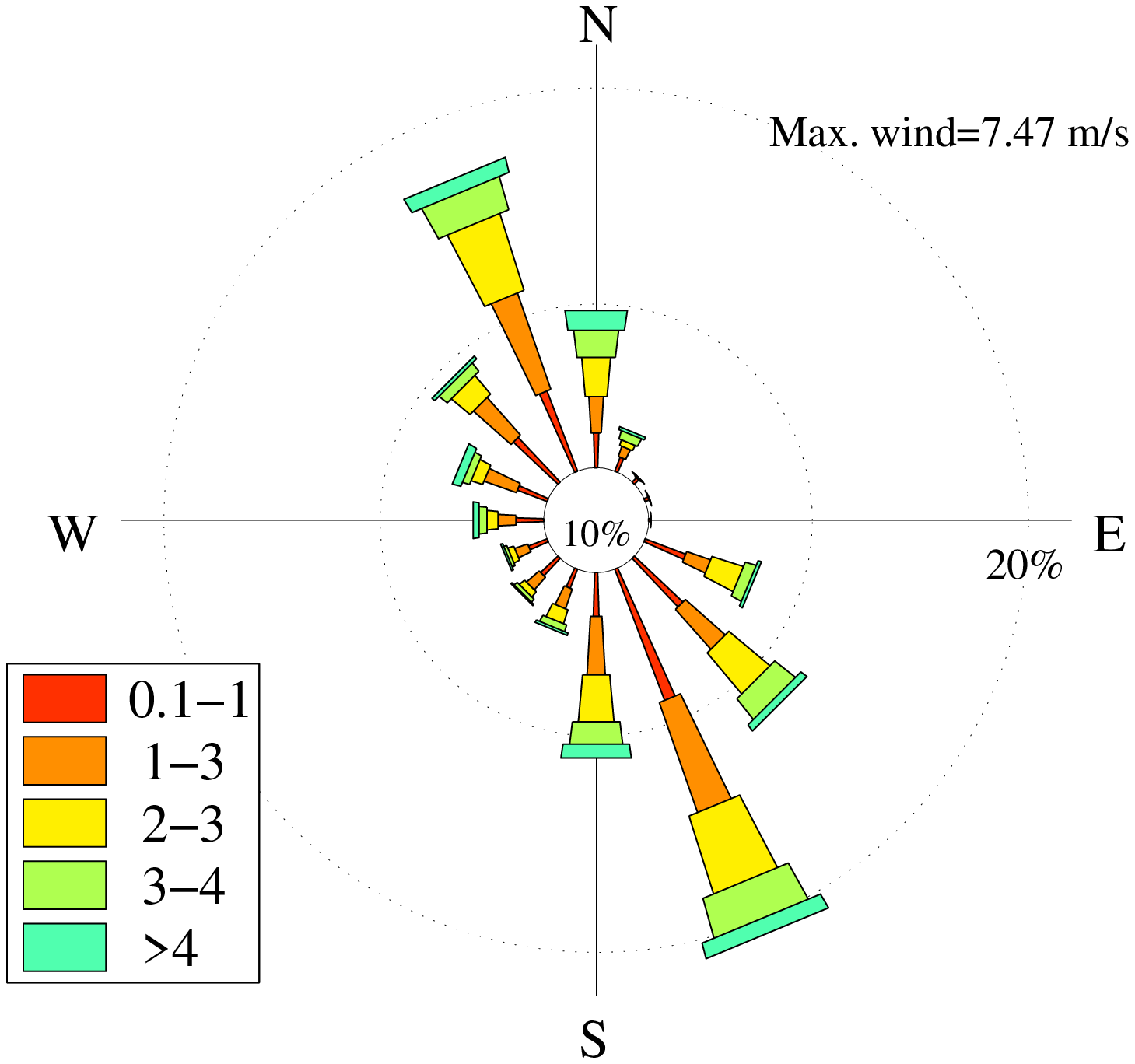} \\
    \includegraphics[width=0.8\columnwidth]{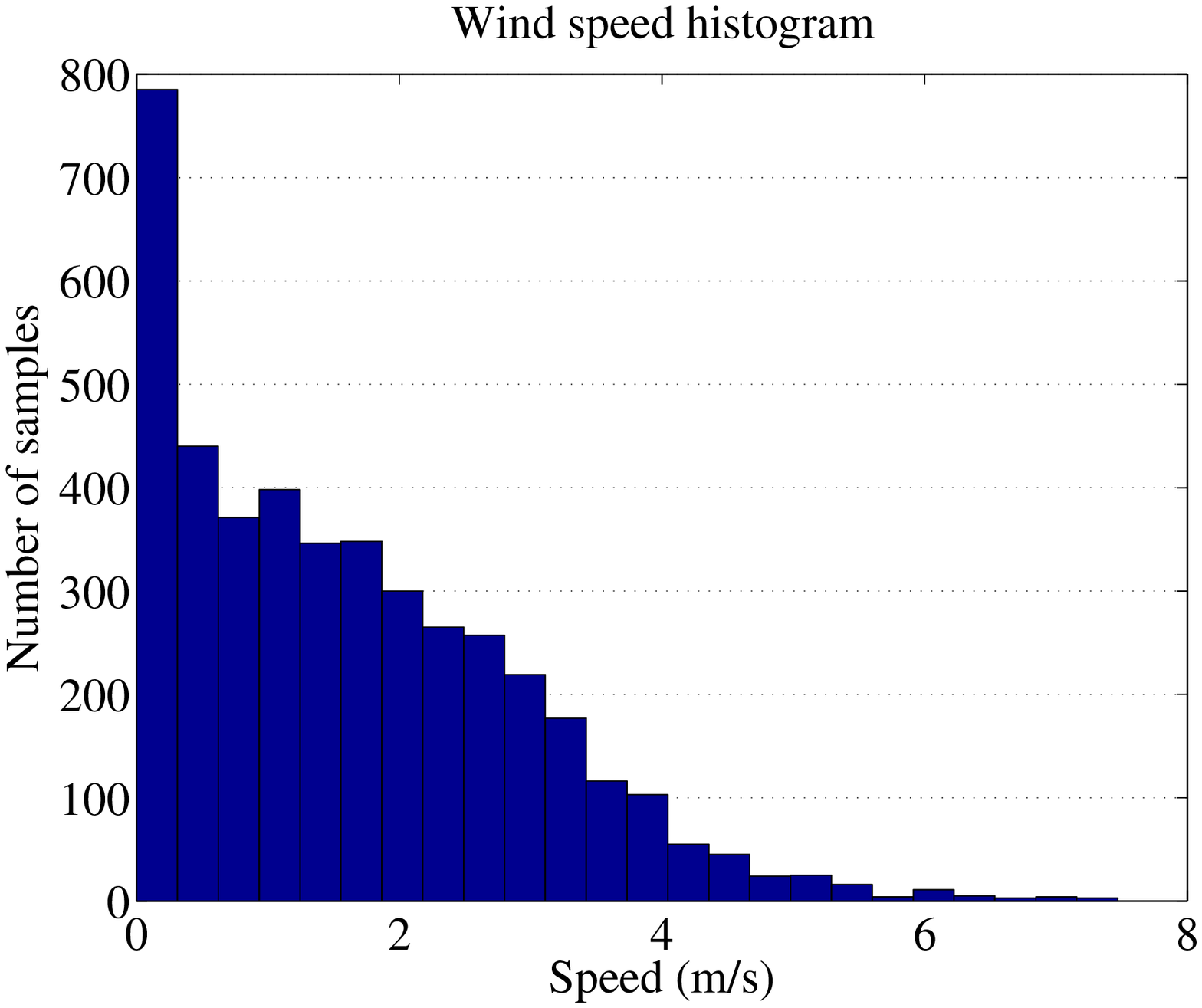}
  }{
    \includegraphics[width=0.8\columnwidth]{runs/save.forward/windrose2}\\
    \includegraphics[width=0.8\columnwidth]{runs/save.forward/windvelh}
  }
  \caption{Measured wind data over the one-month period June 3--July 2,
    2002.  Top: Wind rose diagram, with the proportion of calm winds
    identified in the central circle.  Bottom: Wind speed histogram.}
  \label{fig:winddata}
\end{figure}

The wind data for all simulations come from actual meteorological
measurements taken on site and are specified as point measurements at
10-minute intervals.  Figure~\ref{fig:winddata} depicts the typical
distributions of wind direction and velocity for a representative
one-month period.  The bimodal nature of the wind distribution in a
direction aligned roughly parallel to the Columbia River valley is
evident from the wind rose diagram.
A significant portion of wind measurements exhibit a zero velocity
which cannot be used directly in Eq.~\en{c}.  It is therefore
usual to introduce a ``cut-off'' velocity $\Umin$ such that whenever the
wind satisfies $U\leqslant \Umin$, no contribution is made to the
deposition during that time interval.  We employ a cut-off of
$\Umin=0.1$ which limits the number of neglected time intervals to 10\%
of the total.

\subsection{Sample forward computation}
\label{sec:plume-sample}

\begin{figure}[htbp]
  \centering
  \ifthenelse{\boolean{@IsSubmitted}}{ 
    \includegraphics[width=0.96\columnwidth]{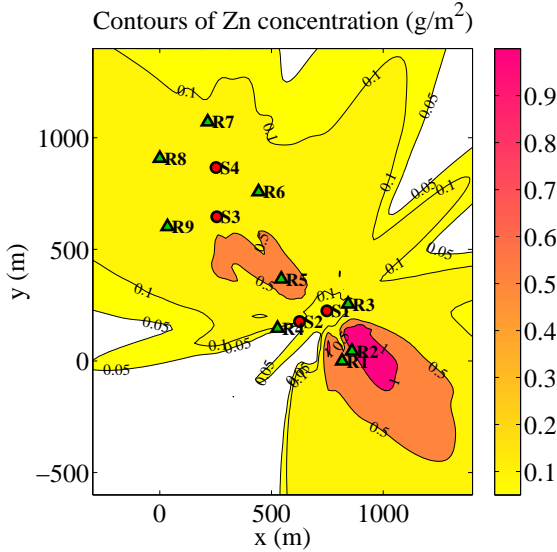}
  }{
    \includegraphics[width=0.96\columnwidth]{runs/save.forward/dep1}
  }
  \caption{Forward simulation of the monthly cumulative deposition of
    zinc (in \units{g \, m^{-2}}) at each point in the domain for source
    emission rates $\bvec{Q}=[35,80,5,5]\times 10^3$ \units{kg\,
      yr^{-1}}.  The wind data used to generate this plot corresponds to
    June 2002, for which the dominant wind direction is towards the
    southeast.  Note that compass north is directed vertically in this
    plot, which is rotated relative to the aerial photo in
    Fig.~{\protect\ref{fig:trailsite}}.} 
  \label{fig:forward1}
\end{figure}

Using the parameter values given in the previous section, we next
calculate a typical distribution of the zinc ground-level concentrations
using the time-varying plume solution~\en{psys-tot}.  The emission rates
for the four sources were taken to be $\bvec{Q}=[35,80,5,5]\times
10^3$~\units{kg\,yr^{-1}}, and measured wind data at 10 minute
intervals was used for the month of June 2002.  The mass of zinc
deposited at each receptor was then computed to 3 significant digits as
\begin{equation}
  \begin{split}
    \bvec{D} = & \left[ \; 17.4,\, 31.0,\, 15.7,\, 1.63,\, 21.7,
    \right.\\
    & \left. \;\; 6.33,\, 2.54,\, 5.02,\, 7.74 \;\right] \times 10^{-6} \;
    \units{kg}.
    \label{eq:depzn}
  \end{split}
\end{equation}
We also computed ground-level zinc concentrations on a 100 $\times$ 100
grid of points covering the entire Trail site, and the results are
displayed as a contour plot in Figure~\ref{fig:forward1}.  The
concentration clearly peaks at locations that lie close to sources S1
and S2 and aligned with the prevailing wind direction.  This is to be
expected since S1 and S2 have by far the largest emission rates among
the four sources.

\section{Inverse problem}
\label{sec:inverse}

We now consider the inverse problem for which values of the zinc
deposition $D_r$ are specified at each receptor $r=1,2,\dots, N_r$, and
the unknowns are the source emission rates $Q_s$ for $s=1,2,\dots, N_s$.
It is convenient to rewrite Eqs.~\en{psys-tot} in the more compact form
\begin{gather}
  \bvec{D} = \mymatrix{P} \, \bvec{Q}
  \label{eq:psys-compact}
\end{gather}
where $\bvec{D}$ and $\bvec{Q}$ are vectors containing the depositions
and emission rates respectively, and $\mymatrix{P}$ is an $N_r\times
N_s$ matrix whose $r,s$ entry is given by
\begin{gather}
  \mymatrixentry{P}_{rs} = \Wdep A \Delta t \, \sum_{n=1}^{N_t}
  p(\bvec{\eta}_r;\bvec{\xi}_s,\theta^n,U^n) |_{z=\Rheight_r} .  
  \label{eq:prs}
\end{gather}
Since there will typically be more receptor measurements than sources
($N_r > N_s$), this is clearly an overdetermined system of equations for
$\bvec{Q}$; hence, there is not a unique solution and we can hope at
best to obtain an approximation of $\bvec{Q}$.

We employ a linear least squares method to determine a solution of
Eq.~\en{psys-compact}, and in all of our computations we employ the {\tt
  lsqlin} solver in MATLAB.  A typical simulation requires only about
20~\units{s} of CPU time on a \computername, and so our method is very
efficient and well-suited to performing parametric studies.
\mycitename{mackay-mckee-mulholland-2006}{MacKay \etal} used a similar
least squares method to estimate parameters such as surface mass
transfer rate and P\'eclet number rather than emission rates.

As a consistency check on our inverse algorithm, we consider emissions
of a single contaminant (zinc) under the influence of wind data during
the month of June 2002.  Deposition values are taken equal to the
outputs from the forward computation in Eq.~\en{depzn},\
with all other parameters chosen the same.  As expected, the emission
rates calculated with the inverse algorithm are identical to within
round-off error.

\subsection{Multiple contaminants and linear constraints}

As mentioned earlier, dustfall measurements are made of several
contaminants, not just zinc.  The contents of each receptor underwent a
trace analysis that measures the mass of a number of trace elements, of
which we are interested in zinc, strontium and sulphur (in the form of
\sulphate).  Although zinc is the contaminant of primary interest in
this study, the sulphur and strontium arise from the same chemical
processes that produce zinc, and consequently we have included them in the
inversion procedure with the hope that they will increase the accuracy
of the zinc estimate.  Because zinc and strontium are deposited as
sulphates, the emission rate of sulphur cannot be included as an
independent variable and so we need to estimate emission rates for
only zinc and strontium.

By modifying our previous notation slightly, we can identify contaminant
species with a superscript $q$ in the emission rates $\bvec{Q}_s^q$
(where $q=$ \zinc\ or \strontium) and depositions $\bvec{D}_r^q$ ($q=$
\zinc, \strontium\ or \sulphate).  Our task is then to solve the
following three linear systems
\begin{align}
  \begin{split}
    \bvec{D}^{\zinc}      &= \mymatrix{P}^{\zinc} \, \bvec{Q}^{\zinc}, \\
    \bvec{D}^{\strontium} &= \mymatrix{P}^{\strontium} \, \bvec{Q}^{\strontium}, \\
    \bvec{D}^{\sulphate}  &= 
    \frac{M^{\sulphate}}{M^{\zinc}} \mymatrix{P}^{\zinc} \bvec{Q}^{\zinc}
    + \frac{M^{\sulphate}}{M^{\strontium}} \mymatrix{P}^{\strontium} \bvec{Q}^{\strontium} 
  \end{split}
  \label{eq:psys2}
\end{align}
where the matrices $\mymatrix{P}^q$ are given in \en{prs} and depend on
the contaminant $q$ through the deposition velocity $\Wdep^q$ and
settling velocity $\Wset^q =g\, \rho^q \, (d^q)^2 /18\mu$.  Since the
scenario we are studying has $N_s=4$ sources and $N_r=9$ receptors,
Eqs.~\en{psys2} represent an overdetermined system of 27 equations in 8
unknowns.  Once $\bvec{Q}^{\zinc}$ and $\bvec{Q}^{\strontium}$ are
known, the emission rate for sulphate can be obtained from
\begin{gather}
  \bvec{Q}^{\sulphate} = \frac{M^{\sulphate}}{M^{\zinc}} \bvec{Q}^{\zinc}
  + \frac{M^{\sulphate}}{M^{\strontium}} \bvec{Q}^{\strontium} .
  \label{eq:qso4}
\end{gather}

In addition to the atmospheric transport processes embodied by the above
equations, the chemical processes generating the contaminants introduce
a number of further constraints on the emission rates.  To be more
specific, we note that source S1 corresponds to a collection of zinc
purification stacks, S2 is a sulphide leach plant cooling tower, and S3
and S4 are identical electrolytic cooling stacks.  As a result, we can
make the following assumptions:
\begin{itemize}
\item Sources 3 and 4 are identical: 
  \begin{gather}
    Q_3^{q}-Q_4^{q}=0, \quad \text{for $q=$ \zinc, \strontium}.
    \label{eq:constraint1}
  \end{gather}

\item The mass ratio of zinc to strontium in sources 3 and 4 is
  approximately 6000 to 1:
  \begin{gather}
    Q_s^{\zinc}-6000 Q_s^{\strontium}=0 \quad \text{for $s=3,4$}.
    \label{eq:constraint2}
  \end{gather}

\item No strontium is emitted from sources 1 and 2:
  \begin{gather}
    Q_1^{\strontium}=Q_2^{\strontium}=0.
    \label{eq:constraint3}
  \end{gather}


\item All emission rates must be non-negative:
  \begin{gather}
    Q_s^{\zinc},\;\; Q_s^{\strontium} \geqslant 0
    \quad \text{for $s=1,2,3,4$}.
    \label{eq:constraint6}
  \end{gather}
\end{itemize}
When taken together, Eqs.~\en{constraint1}--\en{constraint6} represent 6
equality and 8 inequality constraints that supplement the overdetermined
system \en{psys2}.  We can therefore continue to make use of the MATLAB
linear least squares solver {\tt lsqlin} which conveniently permits the
inclusion of both equality and inequality constraints.

\section{Numerical simulations}
\label{sec:sims}

\subsection{Base case}
\label{sec:sims-base}

\begin{figure}
  \centering
  \ifthenelse{\boolean{@IsPstBarFigs}}{ 
    \psset{unit=0.1\columnwidth}%
    \newpsbarstyle{hatcha}{framearc=0,fillstyle=hlines*,hatchangle=45,fillcolor=blue}%
    \newpsbarstyle{hatchb}{framearc=0,fillstyle=hlines*,hatchangle=0,fillcolor=red}%
    \newpsbarstyle{hatchc}{framearc=0,fillstyle=hlines*,hatchangle=135,fillcolor=yellow}%
    \begin{pspicture}(0,-0.5)(9,7.5)%
      \psgrid[xunit=0.9\columnwidth,gridlabels=0,subgriddiv=0,griddots=45](0,0)(1,7)%
      \psaxes[axesstyle=frame,Ox=0,Dx=1,Oy=0,Dy=10,dy=0.1\columnwidth,labels=y,ticks=y](0,0)(9,7)%
      \psbarscale(0.1){}%
      \readpsbardata{\data}{runs/save.base/depall.dat}%
      \psbarchart[chartstyle=cluster,barcolsep=0.2,barstyle={hatcha,hatchb,hatchc}]{\data}%
    \end{pspicture}
  }{
    \ifthenelse{\boolean{@IsSubmitted}}{ 
      \ifthenelse{\boolean{@IsBWplots}}{ 
        \includegraphics[width=0.95\columnwidth]{fig5bw}
      }{
        \includegraphics[width=0.95\columnwidth]{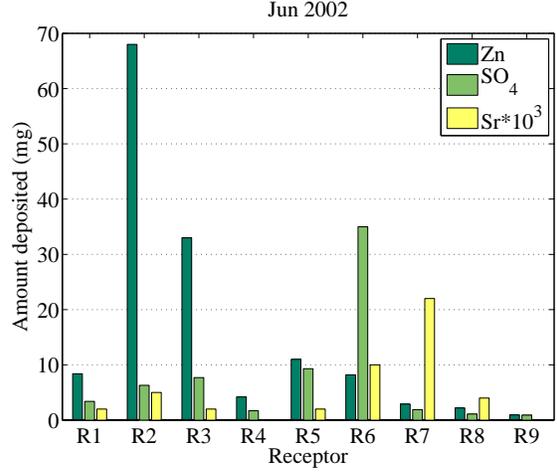}
      }
    }{
      \includegraphics[width=0.95\columnwidth]{runs/save.base/invdepJun02}
    }
  }
  \caption{Experimental measurements of the mass of contaminants
    accumulated within each receptor during the month of June 2002.  The
    zinc and sulphate figures are shown in units of \units{mg}, while
    strontium is scaled by $10^3$ so that it is visible on the same
    axes.}
  \label{fig:dep-jun02}
\end{figure}

\begin{figure}
  \centering
  \ifthenelse{\boolean{@IsPstBarFigs}}{ 
    \psset{unit=0.25\columnwidth}%
    \newpsbarstyle{hatcha}{framearc=0,fillstyle=hlines*,hatchangle=45,fillcolor=blue}%
    \newpsbarstyle{hatchb}{framearc=0,fillstyle=hlines*,hatchangle=0,fillcolor=red}%
    \newpsbarstyle{hatchc}{framearc=0,fillstyle=hlines*,hatchangle=135,fillcolor=yellow}%
    \begin{pspicture}(0,-0.2)(4,3.7)%
      \psgrid[xunit=\columnwidth,gridlabels=0,subgriddiv=0,griddots=45](0,0)(1,3.5)%
      \psaxes[axesstyle=frame,Ox=0,Dx=1,Oy=0,Dy=10,dy=0.1\columnwidth,labels=y,ticks=y](0,0)(4,3.5)%
      \psbarscale(0.1){}%
      \readpsbardata{\datb}{runs/save.base/sources.dat}%
      \psbarchart[chartstyle=cluster,barcolsep=0.2,barstyle={hatcha,hatchb,hatchc}]{\datb}%
    \end{pspicture}
  }{
    \ifthenelse{\boolean{@IsSubmitted}}{ 
      \ifthenelse{\boolean{@IsBWplots}}{ 
        \includegraphics[width=0.95\columnwidth]{fig6bw}
      }{
        \includegraphics[width=0.95\columnwidth]{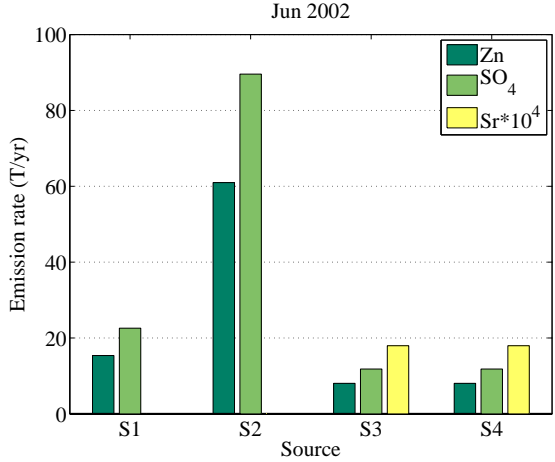}
      }
    }{
      \includegraphics[width=0.95\columnwidth]{runs/save.base/invqrateJun02}
    }
  }
  \caption{Estimated emissions rates for the ``base case'' corresponding
    to June 2002 (the strontium value is scaled by $10^4$).}
  \label{fig:q-jun02}
\end{figure}

We begin by focusing on a number of inverse calculations using a single
month of wind and deposition data corresponding to June 2002 -- we refer
to this simulation as the ``base case.''  The receptor measurements used
as inputs are displayed in Fig.~\ref{fig:dep-jun02}, while the wind data
is the same as that shown earlier in Fig.~\ref{fig:winddata}.
The estimated source emission rates are displayed in
Fig.~\ref{fig:q-jun02} in units of tonnes per year.  We are primarily
interested in zinc emissions, which come to a total of
$92.44\;\units{T\,yr^{-1}}$ for all four sources.  This total should be
compared with the figure of 116.48~\units{T\,yr^{-1}} reported by
Teck-Cominco in the \emph{National Pollutant Release Inventory} for the
year 2002 \mycite{npri}.  The fact that these two figures are so close
gives us some confidence that our inverse approach for estimating
emissions is a reasonable one.  We have not attempted a comparison
between the other two contaminants because strontium emissions are not
reported to NPRI, and our model does not account for all sources of
atmospheric sulphate emissions on the Trail site.

\subsection{Parameter sensitivities}
\label{sec:sims-accuracy}

\begin{figure}
  \centering
  \ifthenelse{\boolean{@IsSubmitted}}{ 
    \ifthenelse{\boolean{@IsBWplots}}{ 
      \includegraphics[width=0.95\columnwidth]{fig7bw}
    }{
      \includegraphics[width=0.95\columnwidth]{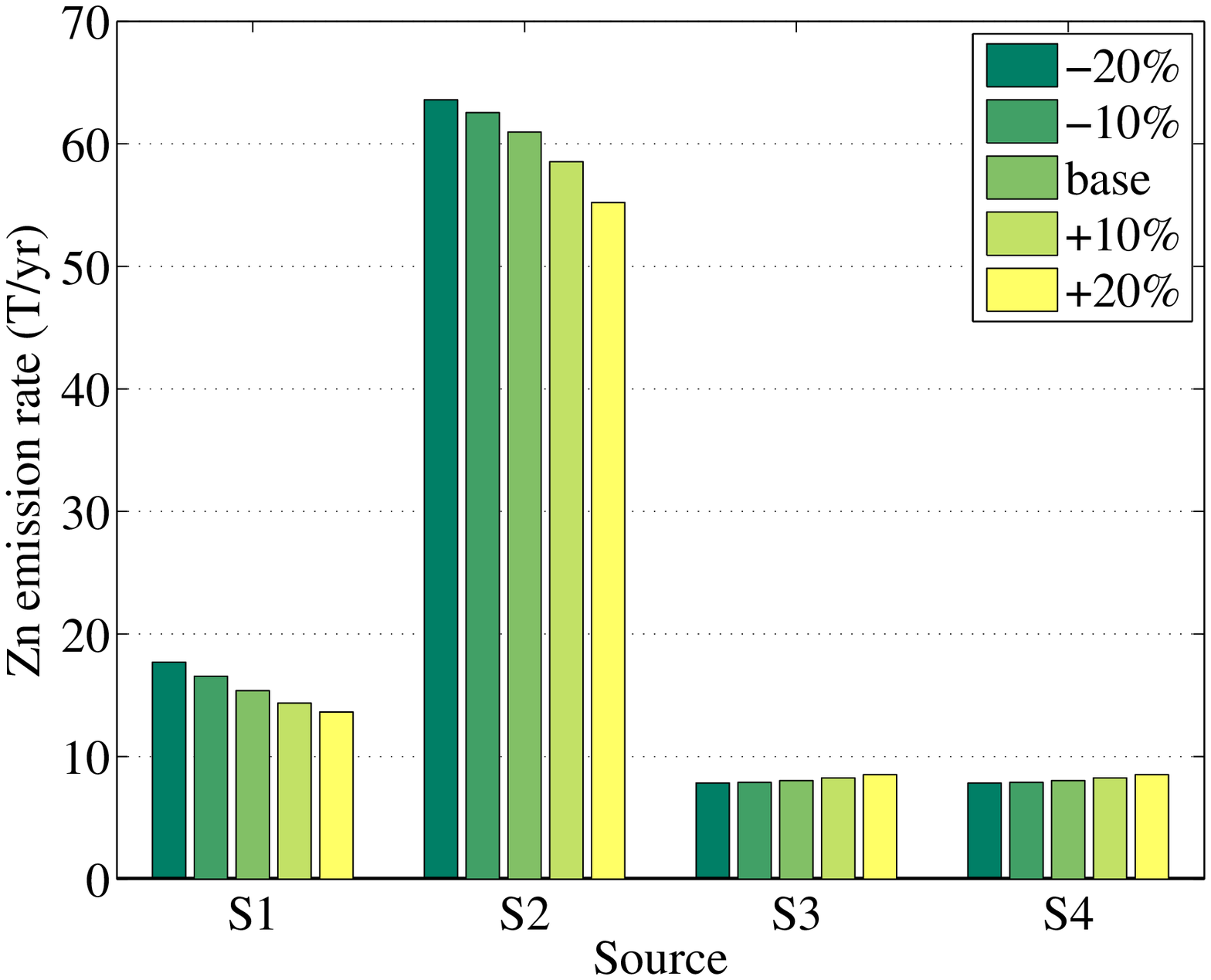}
    }
  }{
    \includegraphics[width=0.95\columnwidth]{runs/summsenshr}
  }
  \caption{Effect of changes in receptor height $\Rheight_r$ on \zinc\  
    emission rates.}
  \label{fig:sens-hr}
\end{figure}

\begin{figure}
  \centering
  \ifthenelse{\boolean{@IsSubmitted}}{ 
    \ifthenelse{\boolean{@IsBWplots}}{ 
      \includegraphics[width=0.95\columnwidth]{fig8bw}
    }{
      \includegraphics[width=0.95\columnwidth]{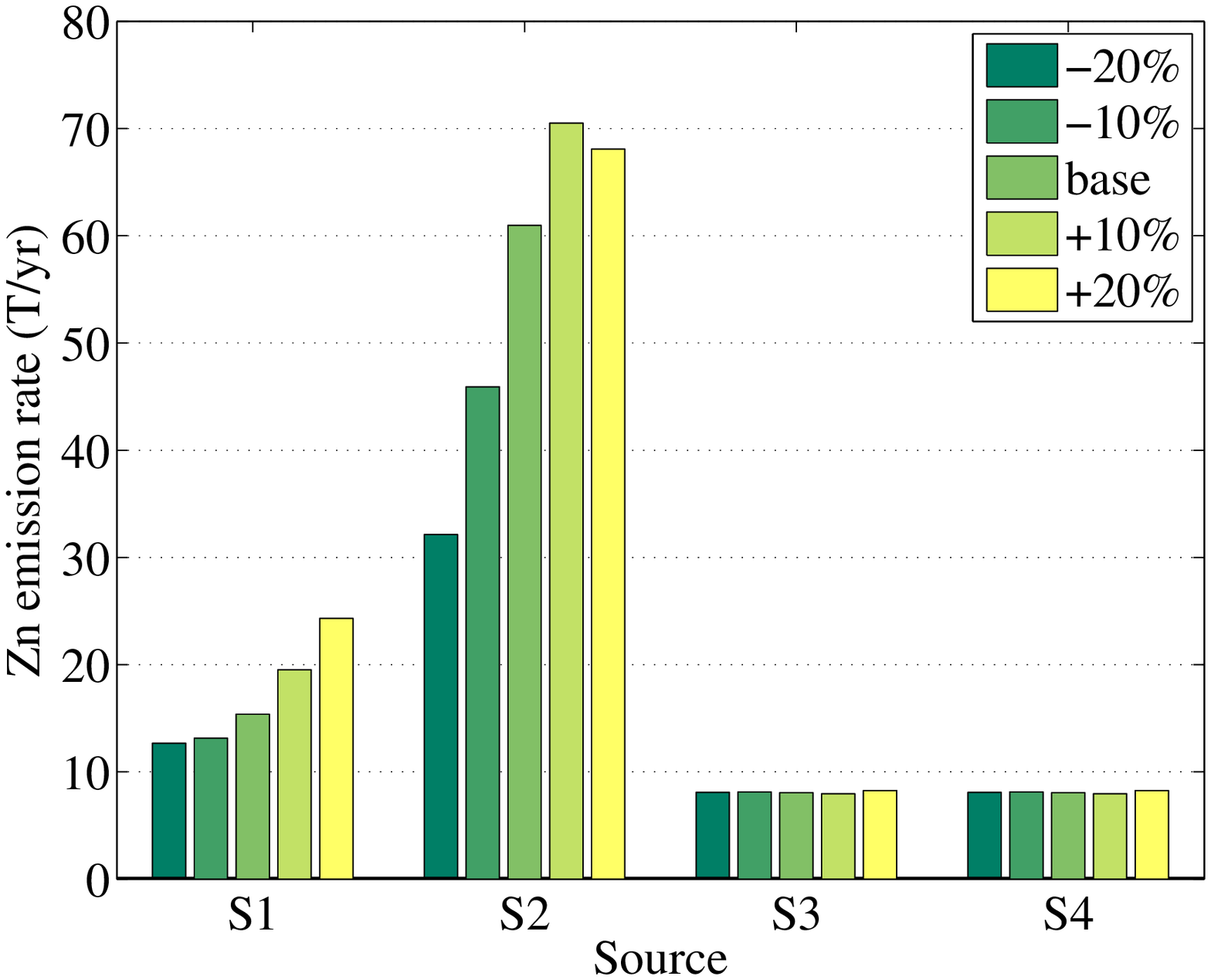}
    }
  }{
    \includegraphics[width=0.95\columnwidth]{runs/summsensHs}
  }
  \caption{Effect of changes in source height $\Sheight_s$ on \zinc\ 
    emission rates.} 
  \label{fig:sens-Hs}
\end{figure}

\begin{figure}
  \centering
  \ifthenelse{\boolean{@IsSubmitted}}{ 
    \ifthenelse{\boolean{@IsBWplots}}{ 
      \includegraphics[width=0.95\columnwidth]{fig9bw}
    }{
      \includegraphics[width=0.95\columnwidth]{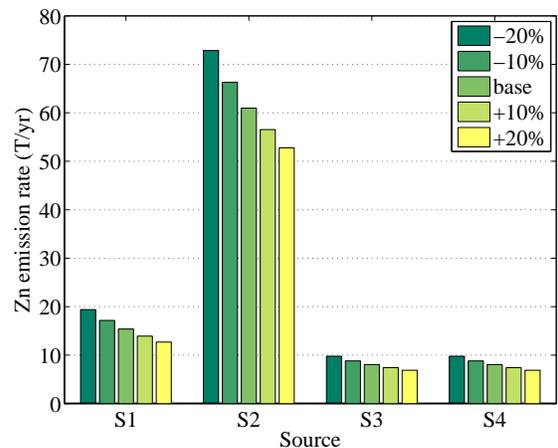}
    }
  }{
    \includegraphics[width=0.95\columnwidth]{runs/summsensWdep}
  }
  \caption{Effect of changes in deposition velocities $\Wdep^q$ on
    \zinc\ emission rates.} 
  \label{fig:sens-Wdep}
\end{figure}

In this section, we investigate the sensitivity of our inverse solution
to changes in several key parameters, namely receptor height
($\Rheight_r$), source height ($\Sheight_s$) and deposition velocity
($\Wdep^q$).  These parameters are singled out because they are all
subject to significant variations for the following reasons:
\begin{itemize}
\item There is a great deal of freedom in the placement of individual
  receptors, which for example can be located at ground level or else on
  the top of buildings or other structure.
\item The height of each source must be adjusted to take into account
  the effect of plume rise, which can be estimated using empirical
  formulas but there remains a significant degree of uncertainty in
  these plume rise estimates.
\item Particle deposition velocities are known to depend strongly on 
  atmospheric conditions and on whether deposition happens under dry or
  wet conditions.
\end{itemize}
We perform a series of simulations, modifying each parameter by $\pm
10$\% and $\pm 20$\% from the base case value.  The resulting emission
estimates are compared in Figs.~\ref{fig:sens-hr}, \ref{fig:sens-Hs} and
\ref{fig:sens-Wdep} for parameters $\Rheight_r$, $\Sheight_s$ and
$\Wdep^q$ respectively.  Only zinc emission rates are depicted here
since similar levels of sensitivity are experienced for the other
contaminants.

From these results, we conclude that the solution is most sensitive to
source height, where relative changes on the order of 40--50\%\ are
seen.  Much less variation is seen in response to changes in $\Sheight$
and $\Wdep$.  These sensitivities are consistent with
\mycitename{miller-hively-1987}{Miller and Hively}, who reviewed a wide
range of Gaussian plume type models and found that for elevated sources,
ground-level concentrations are typically estimated to within a factor
of 0.65 to 1.35 of the actual values.

\subsection{Noise in deposition measurements}
\label{sec:sims-noise}

Even though we expect that the emission rates should remain
approximately constant over time (at least during a given year) there is
nonetheless a significant spread in deposition measurements from month
to month.  There are a number of possible explanations for this
variation, including measurement errors, wet deposition during rainy
periods (where ``scrubbing'' can significantly reduce the amount
deposited) or contamination from secondary sources not accounted for in
the model.  Indeed, \mycitename{goodarzi-etal-2002b}{Goodarzi \etal}
performed an experimental study of emissions at the Teck Cominco's Trail
operation and determined that secondary sources on the site (such as ore
concentrate and slag storage piles) experience resuspension of particles
which may interfere with deposition measurements.

\begin{figure}
  \centering
  \ifthenelse{\boolean{@IsSubmitted}}{ 
    \ifthenelse{\boolean{@IsBWplots}}{ 
      \includegraphics[width=0.95\columnwidth]{fig10bw}
    }{
      \includegraphics[width=0.95\columnwidth]{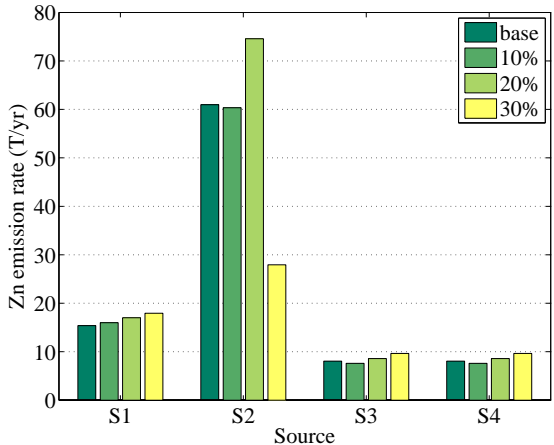}
    }
  }{
    \includegraphics[width=0.95\columnwidth]{runs/summnoise}
  }
  \caption{Effect of random noise in deposition measurements on \zinc\
    emission rates.} 
  \label{fig:noise}
\end{figure}

In any case, it is important to understand the effect that possible
errors in particulate measurements may have on emission estimates.  To
this end, we take each deposition measurement and scale it by a normally
distributed random number chosen from the interval $[1-\alpha,
1+\alpha]$ for values of $\alpha=0.1$, 0.2 and 0.3.  The results are
summarized in Fig.~\ref{fig:noise}, from which it is clear that even for
the largest noise ratio $\alpha=0.3$, the corresponding errors in the
computed emission rates are relatively small.  One possible explanation
for this limited sensitivity to noise is that positive and negative
contributions to errors in the input data may cancel each other out on
average.

\subsection{A note on ill-conditioning}
\label{sec:sims-illcond}

Our approach does not suffer from the extreme levels of sensitivity
observed by others in inverse computations based on Gaussian plume type
solutions (for example, \mycite{miller-hively-1987},
\mycite{enting-newsam-1990a} and \mycite{enting-newsam-1990b}).  We can
explain this apparent discrepancy by noting that our simulations are
performed over relatively short time periods as well as being restricted
to ground level and to areas very close to the emission source.  In
contrast, Enting and Newsam's study of the inverse emissions problem
indicated that computed concentrations at distances further than
10~\units{km} downwind from a source are highly sensitive when
high-altitude pollutant measurements are used.  Although these and other
studies of atmospheric dispersion focus on long-range transport (over
10's or even 100's of \units{km}) our approach benefits from the fact
that we consider transport over much shorter spatial scales.

\subsection{Deposition over several months}
\label{sec:sims-mult}

\begin{figure}
  \centering
  \ifthenelse{\boolean{@IsSubmitted}}{ 
    \includegraphics[width=0.95\columnwidth]{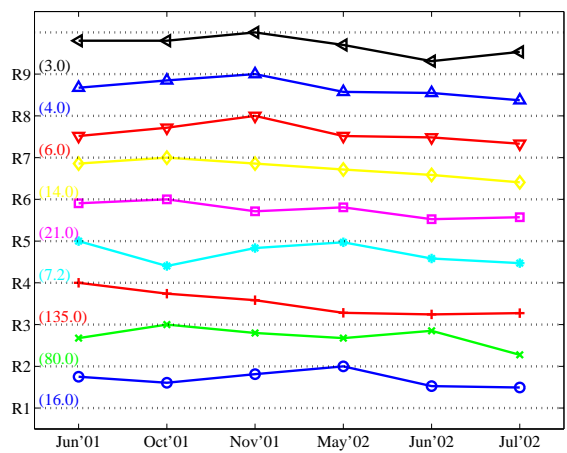}
  }{
    \includegraphics[width=0.95\columnwidth]{runs/depplot4}
  }
  \caption{Simulated \zinc\ deposition rates for every monthly
    period under consideration.  For each receptor, the axis limits are
    drawn as horizontal dotted lines and are scaled so that the lower
    limit is zero and the upper limit represents the maximum deposition
    rate (given by the number in parentheses to the left of each curve
    in \units{T\, yr^{-1}}).}
  \label{fig:depall}
\end{figure}

Deposition measurements are available for a total of six monthly
periods: June, October and November in 2001; and May, June and July in
2002.  All of these measurements correspond to periods during which the
Trail smelter was in continuous operation without being shut down;
therefore, one would expect the depositions to be fairly consistent from
month to month.  The measured values for zinc are summarized in
Fig.~\ref{fig:depall}, from which we observe that there are significant
variations at each receptor location over time, although the trend is
fairly consistent between one receptor and another.  Furthermore, the
largest accumulations are consistently measured in receptors located
closest to the primary sources S1 and S2, as expected.  Similar trends
are observed in the strontium and sulphate data and so they are not
depicted.

\begin{figure}
  \centering
  \ifthenelse{\boolean{@IsSubmitted}}{ 
    \ifthenelse{\boolean{@IsBWplots}}{ 
      \includegraphics[width=0.95\columnwidth]{fig12bw}
    }{
      \includegraphics[width=0.95\columnwidth]{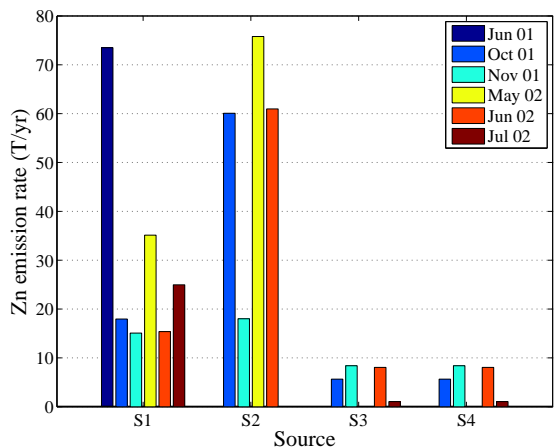}
    }
  }{
    \includegraphics[width=0.95\columnwidth]{runs/summallZn}
  }
  \caption{Computed \zinc\ emission rates for each monthly period.}
  \label{fig:source-zn}
\end{figure}

Keeping in mind our earlier discussion in Section~\ref{sec:sims-noise}
of the solution sensitivity to the error in deposition data, we next
apply our inverse algorithm for each one-month
periods individually.  The resulting monthly emission rate estimates for
zinc are given in the bar plot in Fig.~\ref{fig:source-zn}.\ \ 
There is considerable variation between the results for individual
months, which is not surprising considering the variations in the input
data.  Furthermore, two of the S2 estimates (corresponding to June 2001
and July 2002) are identically zero, indicating that the least squares
algorithm has forced the S2 value up against the boundary of the
inequality constraint $Q_3^{\zinc}\geqslant 0$.  These two results
are clearly questionable because S2 is by far the largest source of
zinc on the site.

With an aim to explaining this discrepancy, we take a closer look at the
zinc deposition data in Fig.~\ref{fig:depall} and observe that the R3
deposition measurements are unusually high in relation to other nearby
receptors (in particular, see the peak value of 135\;\units{mg} measured
in June 2001).  Instead, one would expect that the values at R3 and R4
are much closer to each other since their location relative to the
primary sources S1 and S2 is so similar.  Furthermore, it seems
reasonable that the maximum deposition should occur at a point lying to
the southeast or northwest of sources S1 and S2, rather than at R3 which is
perpendicular to the direction of the prevailing winds.  We conclude
therefore that the large deposition at R3 is most likely attributable to
measurement errors or some other anomaly, and hence R3 should be
excluded from the calculations.

\begin{figure}
  \centering
  \ifthenelse{\boolean{@IsSubmitted}}{ 
    \ifthenelse{\boolean{@IsBWplots}}{ 
      \includegraphics[width=0.95\columnwidth]{fig13bw}
    }{
      \includegraphics[width=0.95\columnwidth]{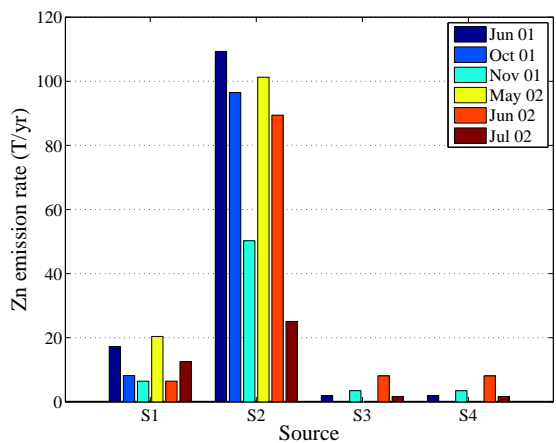}
    }
  }{
    \includegraphics[width=0.95\columnwidth]{runs/summallZnR3}
  }
  \caption{Computed \zinc\ emission rates for each monthly period,
    leaving out receptor R3 (compare to
    Fig.~{\protect\ref{fig:source-zn}}).} 
  \label{fig:source-zn-noR3}
\end{figure}

\begin{figure}
  \centering
  \ifthenelse{\boolean{@IsSubmitted}}{ 
    \includegraphics[width=0.95\columnwidth]{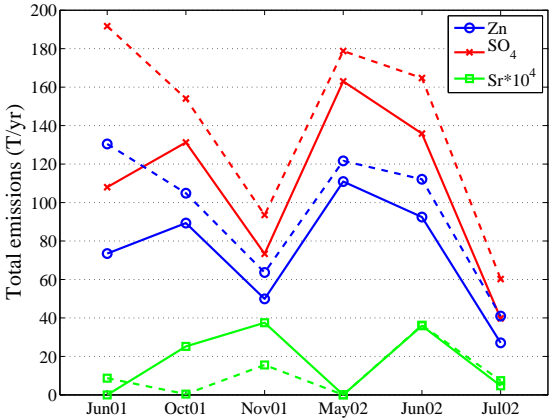}
  }{
    \includegraphics[width=0.95\columnwidth]{runs/sourcetotboth}
  }
  \caption{Total emissions for each contaminant.  The solid lines
    correspond to simulations with R3 included in the inverse
    calculation, while the dashed lines are without R3.}
  \label{fig:sourcetot}
\end{figure}

Upon repeating the previous set of simulations with R3 excluded, we
obtain the results shown in Fig.~\ref{fig:source-zn-noR3}.  All S1 and
S2 emission estimates are now nonzero and the anomalously high S1
estimate for June 2002 is reduced to be in more line with the other
values.  Although there remain significant variations between a few of
the monthly estimates (namely November 2001 and July 2002), these
results are much more reasonable.
Since the primary quantity of interest in this study is the total
emissions from all sources, another way of viewing these results is via
the total emissions from all four sources for each monthly period, as
shown in Fig.~\ref{fig:sourcetot}. 

If we assume that emissions are constant throughout the year, then we
may also use all months of deposition data from a given year to estimate
a single ``aggregate'' emission rate.  The aggregate estimates based on
the 2001 and 2002 data are shown in Table~\ref{tab:totals} along with
the total emission rates for the individual months.  One advantage of
using the aggregate calculation is that it serves to average out some of
the variation between the individual monthly results.  The aggregate
figures can then be compared to the values from the NPRI database of
100.33~\units{T\,yr^{-1}} in 2001 and 116.48~\units{T\,yr^{-1}} in
2002.  For both years, the aggregate estimates are within 10-20\%\ of
the reported values, and an increasing trend from 2001 to 2002 is also
captured.
\begin{table}
  \centering
  \caption{Summary of total zinc emissions for each month of interest,
    the corresponding aggregate totals for each year, and the emissions
    reported to NPRI.  Receptor R3 is omitted.} 
  \label{tab:totals}
  \begin{tabular}{|l|cc|}\hline
              & \multicolumn{2}{c|}{\mbox{\;} \zinc\ Emission rates 
                (\units{T\,yr^{-1}})\mbox{\;}}\\  
              & \quad\; 2001 \quad\qquad  & \quad 2002 \;\qquad \\ \hline\hline
    \multirow{3}{2.6cm}{\mbox{\;}Monthly estimates}
              & 130.5                     & 121.7 \\
              & 104.9                     & 112.1 \\
              &  63.64                    &  40.97 \\\hline
    \;Aggregate estimates
              &  95.77                    & 104.5 \\\hline
    \;Reported to NPRI{\protect\mycite{npri}}\;
              & 100.33                    & 116.48\\\hline
  \end{tabular}
\end{table}

\section{Conclusions}
\label{sec:conclude}

We have derived a linear least squares approach for estimating
contaminant emissions from several point sources given time-varying wind
data and monthly-averaged deposition measurements.  The model is based
on Ermak's Gaussian plume type solution to the advection-diffusion
equations~\mycite{ermak-1977} which incorporates particle settling and
surface deposition.  The novelty of our approach stems from its focus on
short range emissions (within 1000~\units{m} of the source),
incorporating additional equality and inequality constraints on
contaminant sources, and its ability to handle time-varying wind data.

The algorithm is used to estimate emissions of several contaminants from
a smelting operation in Trail, British Columbia using dustfall
measurements taken over several one-month periods during 2001--2002.  We
demonstrate that the algorithm is efficient and robust to changes in
several key parameter values in comparison to other approaches published
in the literature.  While there is significant variation from
month-to-month and between the individual sources, the average annual
estimate of the emission rate is within 10--20\%\ of the values reported to
Environment Canada for the years in question.  

There remain a number of aspects of the model which require further
study and refinement, for example by using more careful estimates of the
contribution of plume rise to source heights, incorporating the effects
of wet deposition during rainy periods, and allowing further variation
in atmospheric conditions through the Pasquill-Gifford stability
classification.  We would also like to apply our insight into
contaminant plume dynamics to guide the design of future studies of the
Trail site; in particular, by relocating a receptor such as R3 to a more
advantageous position (e.g., within the areas of peak concentration
southeast or northwest of sources S1 and S2).  Finally, we will perform
a direct finite volume discretization of the 3D advection-diffusion
equation, for which reliable estimates for the turbulent eddy
diffusivities are essential.  Comparisons of these simulations with the
results of our inverse Gaussian plume algorithm will appear in a companion
publication~\cite{lebed-stockie-2009}.

\leavethisout{
  \begin{itemize}
  \item A few more details to address:
    \begin{itemize}
    \item limit of calm winds, 
    \item steady state assumption,
    \item spatial variation in the eddy diffusivities.  See
      \mycitename{lange-1978}{Lange} uses a height-dependent diffusion
      coefficient with $K_z(z)=0.1 z$ for $0\leqslant z \leqslant 100\;m$.  Others
      that give similar expressions for the vertical variation are
      \mycitename{koch-1989}{Koch},
      \mycitename{lettau-dabberdt-1970}{Lettau and Dabberdt},
      \mycitename{obrien-1970}{O'Brien},
      \mycitename{yadav-etal-2003}{Yadav \etal}.
      \mycitename{lin-hildemann-1997}{Lin and Hildemann} have analytical
      solutions to compare to.
    \end{itemize}
  \end{itemize}
}

\appendix
\ack

We are indebted to Ed Kniel of Teck Cominco Limited for providing us
with experimental data and also for many insightful discussions.
Funding for this research was provided by the Natural Sciences and
Engineering Research Council of Canada (NSERC), the MITACS Network of
Centres of Excellence, and Teck Cominco.  JMS acknowledges the support
of the Alexander von~Humboldt Foundation and Fraunhofer Institut f\"ur
Techno- und Wirtschaftsmathematik in Kaiserslautern.



\ifthenelse{\boolean{@IsSubmitted}}{

}{
  \bibliographystyle{elsart-num}
  \bibliography{long-abbrevs,atmos,mybooks,numanal,stockie}

\providecommand{\noopsort}[1]{}
\begin{thebibliography}{10}
\expandafter\ifx\csname url\endcsname\relax
  \def\url#1{\texttt{#1}}\fi
\expandafter\ifx\csname urlprefix\endcsname\relax\def\urlprefix{URL }\fi

\bibitem{turner-1979}
D.~B. Turner, Atmospheric dispersion modeling: {A} critical review, Journal of
  the Air Pollution Control Association 29~(5) (1979) 502--519.

\bibitem{settles-2006}
G.~S. Settles, Fluid mechanics and homeland security, Annual Review of Fluid
  Mechanics 38 (2006) 87--110.

\bibitem{richardson-1921}
L.~F. Richardson, Some measurements of atmospheric turbulence, Philosophical
  Transactions of the Royal Society of London A 221~(1921) (1920) 1--28.

\bibitem{taylor-1922}
G.~I. Taylor, Diffusion by continuous movements, Proceedings of the London
  Mathematical Society (1922) 196--212.

\bibitem{sutton-1932}
O.~G. Sutton, A theory of eddy diffusion in the atmosphere, Proceedings of the
  Royal Society of London, Series A 135~(826) (1932) 143--165.

\bibitem{ermak-1977}
D.~L. Ermak, An analytical model for air pollutant transport and deposition
  from a point source, Atmospheric Environment 11~(3) (1977) 231--237.

\bibitem{calder-1977}
K.~L. Calder, Multiple-source plume models of urban air pollution -- {T}heir
  general structure, Atmospheric Environment 11 (1977) 403--414.

\bibitem{liley-1995}
J.~B. Liley, Analytic solution of a one-dimensional equation for aerosol and
  gas dispersion in the stratosphere, Journal of the Atmospheric Sciences
  52~(18) (1995) 3283--3288.

\bibitem{lin-hildemann-1997}
J.-S. Lin, L.~M. Hildemann, A gen\-er\-al\-ized math\-e\-mat\-ic\-al scheme to
  analytically solve the atmospheric diffusion equation with dry deposition,
  Atmospheric Environment 31~(1) (1997) 59--71.

\bibitem{condie-bormans-1997}
S.~A. Condie, M.~Bormans, The influence of density stratification on particle
  settling, dispersion and population growth, Journal of Theoretical Biology
  187 (1997) 65--75.

\bibitem{elbadia-etal-2005}
A.~{El Badia}, T.~Ha-Duong, A.~Hamdi, Identification of a point source in a
  linear advection-dispersion-reaction equation: application to a pollution
  source problem, Inverse Problems 21 (2005) 1121--1136.

\bibitem{kennedy-ericsson-wong-2005}
C.~Kennedy, H.~Ericsson, P.~L.~R. Wong, Gaussian plume modeling of contaminant
  transport, Stochastic and Environmental Risk Assessment 20 (2005) 119--125.

\bibitem{jeong-etal-1995}
H.-J. Jeong, E.-H. Kim, K.-S. Suh, W.-T. Hwang, M.-H. Han, H.-K. Lee,
  Determination of the source rate released into the environment from a nuclear
  power plant, Radiation Protection Dosimetry 113~(3) (2005) 308--313.

\bibitem{hogan-etal-2005}
W.~R. Hogan, G.~F. Cooper, M.~M. Wagner, G.~L. Wallstrom, An inverted
  {G}aussian plume model for estimating the location and amount of release of
  airborne agents from downwind atmospheric concentrations, {RODS} technical
  report, Realtime Outbreak and Dis\-ease Sur\-veil\-lance Lab\-o\-ra\-to\-ry,
  University of Pittsburgh, Pittsburgh, PA (2005).

\bibitem{mackay-mckee-mulholland-2006}
C.~MacKay, S.~McKee, A.~J. Mulholland, Diffusion and convection of gaseous and
  fine particulate from a chimney, IMA Journal of Applied Mathematics 71 (2006)
  670--691.

\bibitem{mulholland-seinfeld-1995}
M.~Mulholland, J.~H. Seinfeld, Inverse air pollution modelling of urban-scale
  carbon monoxide emissions, Atmospheric Environment 29~(4) (1995) 497--516.

\bibitem{seibert-frank-2004}
P.~Seibert, A.~Frank, Source-receptor matrix calculation with a {L}agrangian
  particle dispersion model in backward mode, Atmospheric Chemistry and Physics
  4 (2004) 51--63.

\bibitem{enting-2002}
I.~G. Enting, Inverse Problems in Atmospheric Constituent Transport, Cambridge
  Atmospheric and Space Science Series, Cambridge University Press, 2002.

\bibitem{goyal-etal-2005}
A.~Goyal, M.~J. Small, K.~{von Stackelberg}, D.~Burmistrov, N.~Jones,
  Estimation of fugitive lead emission rates from secondary lead facilities
  using hierarchical {B}ayesian models, Environmental Science and Technology
  39~(13) (2005) 4929--4937.

\bibitem{bagtzoglou-baun-2005}
A.~C. Bagtzoglou, S.~A. Baun, Near real-time atmospheric contamination source
  identification by an optimization-based inverse method, Inverse Problems in
  Science and Engineering 13~(3) (2005) 241--259.

\bibitem{brown-1993}
M.~Brown, Deduction of emissions of source gases using an objective inversion
  algorithm and a chemical transport model, Journal of Geophysical Research
  98~(D7) (1993) 12639--12660.

\bibitem{houweling-etal-1999}
S.~Houweling, T.~Kaminski, F.~J. Dentener, J.~Lelieveld, M.~Heimann, Inverse
  modeling of methane sources and sinks using the adjoint of a global tranport
  model, Journal of Geophysical Research 104~(D21) (1999) 26137--26160.

\bibitem{beychok-1999}
M.~R. Beychok, Error propagation in air dispersion modeling, ChemAlliance.org,
  \url{http://www.chemalliance.org/Articles/Industry/Ind990816.asp} (16 August
  1999).

\bibitem{goodarzi-etal-2002b}
F.~Goodarzi, H.~Sanei, M.~Labont\'e, W.~F. Duncan, Sources of lead and zinc
  associated with metal smelting activities in the {T}rail area, {B}ritish
  {C}olumbia, {C}anada, Journal of Environmental Monitoring 4 (2002) 400--407.

\bibitem{npri}
National pollutant release inventory, Environment Canada, {URL}:
  \url{http://www.ec.gc.ca/pdb}, accessed on April 27, 2009 (Trail, British
  Columbia, NPRI ID~\#3802).

\bibitem{hanna-briggs-hosker-1982}
S.~R. Hanna, G.~A. Briggs, R.~P. {Hosker Jr.}, Handbook on atmospheric
  diffusion, Tech. Rep. DOE/TIC-11223, Technical Information Center, U.S.
  Department of Energy (1982).

\bibitem{chrysikopoulos-hildemann-roberts-1992}
C.~V. Chrysikopoulos, L.~M. Hildemann, P.~V. Roberts, A three-dimensional
  steady-state atmospheric dispersion-deposition model for emissions from a
  ground-level area source, Atmospheric Environment 26A~(5) (1992) 747--757.

\bibitem{koch-1989}
W.~Koch, A solution of the two-dimensional atmospheric diffusion equation with
  height-dependent diffusion coefficient including ground level absorption,
  Atmospheric Environment 23~(8) (1989) 1729--1732.

\bibitem{gatz-1975}
D.~F. Gatz, Pollutant aerosol deposition into southern {L}ake {M}ichigan,
  Water, Air and Soil Pollution 5 (1975) 239--251.

\bibitem{mcmahon-denison-1979}
T.~A. McMahon, P.~J. Denison, Empirical atmospheric deposition parameters --
  {A} review, Atmospheric Environment 13 (1979) 571--585.

\bibitem{pacyna-etal-1989}
J.~M. Pacyna, A.~Bartonova, P.~Cornille, W.~Maenhaut, Modelling of long-range
  transport of trace elements. {A} case study, Atmospheric Environment 23~(1)
  (1989) 107--114.

\bibitem{briggs-1973}
G.~A. Briggs, Diffusion estimation for small emissions, Atmospheric Turbulence
  and Diffusion Laboratory Contribution, File No.~79, National Oceanic and
  Atmospheric Administration, Oak Ridge, TN (1973).

\bibitem{carrascal-etal-1993}
M.~D. Carrascal, M.~Puigcerver, P.~Puig, Sensitivity of {G}aussian plume model
  to dispersion specifications, Theoretical and Applied Climatology 48 (1993)
  147--157.

\bibitem{miller-hively-1987}
C.~W. Miller, L.~M. Hively, A review of validation studies for the {G}aussian
  plume atmospheric dispersion model, Nuclear Safety 28~(4) (1987) 522--531.

\bibitem{enting-newsam-1990a}
I.~G. Enting, G.~N. Newsam, Atmospheric constitutent in\-ver\-sion prob\-lems:
  {I}m\-pli\-ca\-tions for base\-line mon\-i\-tor\-ing, Journal of Atmospheric
  Chemistry 11 (1990) 69--87.

\bibitem{enting-newsam-1990b}
I.~G. Enting, G.~N. Newsam, Inverse problems in atmospheric constituent
  studies: {II}. {S}ources in the free atmosphere, Inverse Problems 6 (1990)
  349--362.

\bibitem{lebed-stockie-2009}
E.~Lebed, J.~M. Stockie, A high-resolution finite volume approach for the
  transport of particulate emissions in the atmosphere, In preparation.

\end{thebibliography}
}


\end{document}